\documentclass{ws-ijmpc}

\begin{document}

\markboth{Nuno Crokidakis}
{Modeling the Siege of Syracuse: Resources, strategy, and collapse}

\catchline{}{}{}{}{}

\title{Modeling the Siege of Syracuse: Resources, strategy, and collapse}

\author{Nuno Crokidakis $^{*}$}

\address{
Instituto de F\'{\i}sica, \hspace{1mm} Universidade Federal Fluminense \\
 Niter\'oi - Rio de Janeiro, \hspace{1mm} Brazil \\ 
$^{*}$ nunocrokidakis@id.uff.br}

\maketitle

\begin{history}
\received{Day Month Year}
\revised{Day Month Year}
\end{history}

\begin{abstract}
\noindent
The Siege of Syracuse (214 - 212 BC) was a decisive event in the Second Punic War, leading to the city's fall to Rome despite its formidable defenses, including the war machines devised by Archimedes. In this work, we propose a mathematical model to describe the dynamics of the siege, incorporating the depletion of resources, the decline of Syracuse's population, and the persistence of the Roman army. Our analysis reveals the existence of a critical threshold $\lambda_c$, which determines the outcome of the siege. This threshold marks a phase transition: if the effectiveness of Syracuse's defenses, represented by $\lambda$, had exceeded $\lambda_c$, the city could have withstood the Roman assault. However, since history records Syracuse's fall, we conclude that $\lambda < \lambda_c$. This result provides a quantitative framework to understand the inevitability of the city's conquest and demonstrates how mathematical modeling can offer new insights into historical military conflicts. We also explore different scenarios and assess the impact of key factors such as siege duration, supply constraints, and defensive capabilities. The results provide insights into how strategic elements influenced the eventual fall of Syracuse and demonstrate the applicability of mathematical modeling in historical military analysis.

\keywords{Dynamics of social systems, Social conflicts, Imperial dynamics, Siege of Syracuse, Mathematical modeling, Resource depletion, Second Punic War, Military strategy, Historical analysis, Phase transitions}

\end{abstract}

\ccode{PACS Nos.: 05.10.-a, 05.70.Jk, 87.23.Ge, 89.75.Fb}

\section{Introduction}

\qquad The siege of Syracuse (214 - 212 BC) was a pivotal event during the Second Punic War, in which the Roman Republic besieged and ultimately conquered the powerful Greek city-state of Syracuse, located in Sicily. This siege marked a crucial moment in Rome's expansionist ambitions and solidified its control over the Mediterranean. Syracuse, previously an ally of Rome under King Hiero II, shifted its allegiance to Carthage after Hiero's death, prompting a Roman military response under the command of Marcus Claudius Marcellus \cite{livy}.

The city's defenses were formidable, bolstered by the engineering ingenuity of Archimedes, whose war machines significantly delayed the Roman assault \cite{plutarch}. Despite these defenses, the prolonged siege led to resource depletion, attrition, and eventual Roman victory through a combination of blockade, direct assault, and internal betrayal. Historical accounts suggest that the Roman besieging force numbered around 20,000–35,000 soldiers, while Syracuse housed a population of approximately 75,000, including defenders and civilians \cite{goldsworthy_2001}.

The siege of Syracuse has been recognized not only in historical analyses but also in popular culture. A notable example is its depiction in the film \textit{Indiana Jones and the Dial of Destiny} \cite{indi_movie}, where the event serves as a significant historical backdrop. The movie portrays the siege as a pivotal moment, emphasizing the strategic importance of the city and the ingenuity of its defenses, particularly the legendary contributions of Archimedes. While the film takes creative liberties, its inclusion of the siege underscores its enduring historical significance and how it continues to capture the imagination of modern audiences.

Integrating references from popular culture into historical analyses can help bridge academic discourse with broader public interest. In the case of the Siege of Syracuse, the cinematic portrayal brings attention to key historical themes such as military strategy, technological innovation, and the broader geopolitical struggles of the ancient Mediterranean. Although the film does not provide a rigorous historical reconstruction, its narrative highlights the same fundamental elements explored in this study: the dynamics of defense, resource management, and the ultimate fall of the city under Roman persistence.

The historical accounts of the siege are well-documented in ancient sources, including Polibius \cite{Polibius}, Livy \cite{livy}, and Plutarch \cite{plutarch}, who describe the military strategies employed and the significant role played by Archimedes in the city's defense. Modern historiography also explores the siege's implications in the broader context of the Punic Wars, as discussed by Goldsworthy \cite{goldsworthy_2001} and Lazenby \cite{Lazenby}, emphasizing its strategic consequences for Rome's dominance over the Mediterranean.

Mathematical modeling has been increasingly employed to analyze historical events, particularly in the study of warfare and societal dynamics \cite{galam_book,csf,sooknanan}. One of the earliest and most influential approaches is Lanchester’s laws, which describe the mathematical relationship between opposing military forces in battle \cite{lanchester1,lanchester2}. These laws distinguish between direct one-on-one combat, where force strength is proportional to army size, and modern ranged warfare, where force effectiveness scales quadratically with army size. Another pioneering work is that of Lewis Fry Richardson, who developed differential equation models to study the dynamics of arms races and the statistical patterns underlying war frequency and intensity \cite{Richardson1,Richardson2}. More broadly, the field of cliodynamics has emerged as a quantitative framework for analyzing historical processes, integrating mathematical models with historical data to study the rise and fall of civilizations \cite{turchin1,turchin2}. These approaches provide a foundation for modeling historical sieges, such as the Siege of Syracuse, where resource depletion, military engagement, and strategic decisions played a crucial role in determining the outcome.

Related approaches have also been proposed to model the historical dynamics of complex societies. For instance, Gunduz \cite{gunduz_empire} developed a mathematical model to describe the rise and fall of empires, using a system of nonlinear differential equations to capture the evolution of political power over time. This type of framework reinforces the relevance of applying simplified mathematical models to analyze the structural dynamics behind major historical transitions, such as the fall of Syracuse.

Given the prolonged nature of the siege and its strategic importance, mathematical modeling offers a valuable framework for analyzing the dynamics of resource depletion, population attrition, and military engagement over time. This study employs a logistic depletion model to examine the impact of siege warfare on a walled city's ability to sustain itself under prolonged military pressure. By incorporating factors such as food supplies, combat losses, and defensive capabilities, we aim to provide a quantitative perspective on the factors leading to the fall of Syracuse.

The fall of Syracuse can be understood in terms of a critical threshold in the effectiveness of its defenses. In the mathematical model developed in this work, we identify a critical value $\lambda_c$ that marks a transition between two possible outcomes: the survival of Syracuse or its eventual conquest by Rome. This behavior resembles a phase transition in physics, where small variations in parameters lead to qualitatively different macroscopic states. If the defensive effectiveness $\lambda$ had exceeded the critical threshold $\lambda_c$, Syracuse might have been able to resist the Roman siege indefinitely. However, since historical records indicate the city's fall, we conclude that $\lambda < \lambda_c$, confirming that the defenses, despite their ingenuity - including the war machines attributed to Archimedes - were ultimately insufficient to prevent Roman victory. This approach provides a novel perspective on the siege, connecting historical events to universal principles of dynamical systems and critical phenomena.

Thus, in this work, we develop a system of three differential equations to model the interactions among available resources, the besieged population, and the Roman attacking force. These equations capture the depletion of resources over time, the decline in the defending population due to starvation and combat, and the attrition of the Roman army. By solving these equations numerically, we explore different scenarios and assess the conditions under which Syracuse ultimately fell.


\section{Mathematical model}

\qquad To analyze the dynamics of the Siege of Syracuse, we develop a system of three differential equations representing the interaction among available resources, the besieged population, and the attacking Roman forces. The model captures key factors such as resource depletion, attrition due to starvation and combat, and the effectiveness of defensive and offensive strategies.

We define the following variables: $R(t)$ is the amount of available resources inside the city at time $t$. In addition, $P(t)$ is the population of Syracuse, including both civilians and defenders, at time $t$. Finally, $A(t)$ is the size of the attacking Roman army at time $t$. The system of differential equations governing the dynamics is given by:
\begin{eqnarray} \label{eq1}
\frac{dR}{dt} & = & -\alpha R + \beta \\ \label{eq2}
\frac{dP}{dt} & = & -\gamma P + \delta R - \kappa P A \\ \label{eq3}
\frac{dA}{dt} & = & + \mu A -\lambda A P 
\end{eqnarray}

Let us elaborate about the parameters:

\begin{itemize}

\item  $\alpha$ (resource depletion rate): Represents how quickly resources inside the city are consumed. A higher value means a faster exhaustion of supplies, leading to a shorter period before famine sets in.

\item $\beta$ (resource replenishment rate): Accounts for any potential inflow of supplies, such as smuggled goods or hidden reserves. If $\beta=0$, the city is fully blockaded, and resources only decline.

\item $\gamma$ (natural mortality rate): Represents the fraction of the population lost due to non-military factors such as disease, malnutrition, and natural causes. A higher $\gamma$ reflects worsening conditions within the city.

\item $\delta$ (resource impact on survival): Captures the beneficial effects of resources on population maintenance. A larger $\delta$ means that having more resources significantly improves survival rates.

\item $\kappa$ (attack-induced population loss): Represents the rate at which the population decreases due to direct attacks by the Roman army. A higher $\kappa$  suggests more aggressive and lethal military engagements against the defenders.

\item $\mu$ (Roman army reinforcement): Represents the rate at which the Roman army is replenished. This replenishment could come from multiple sources, including reinforcements arriving by sea, recruitment of local mercenaries, or supply lines from Roman-controlled territories.

\item $\lambda$ (effectiveness of Syracuse's defenses): Represents the ability of the city's defenders to inflict losses on the Roman army. This includes all forms of defensive strategies, such as fortifications, military resistance, and the war machines designed by Archimedes. A higher $\lambda$ indicates a stronger defensive effort, making it more difficult for the Romans to sustain their siege.

\end{itemize}

The first equation models the decline of resources due to consumption by the population and natural degradation over time. The second equation captures population losses due to natural causes, starvation, and direct military confrontations, but it presents a term that models the survivance due to limited resources. The third equation represents the attrition of the Roman army caused by resistance from the defenders, and Roman army reinforcement. It is important to note that the model does not include non-combat-related deaths of Roman soldiers, such as those resulting from disease, malnutrition, or exhaustion. This simplification was made to preserve analytical tractability and to focus on the primary mechanisms driving the outcome of the siege.

This system provides a framework for analyzing how the balance between resources, defensive effectiveness, and Roman military pressure influenced the duration and outcome of the siege. Numerical simulations of this model allow for the exploration of different scenarios and the identification of key parameters that determined the fall of Syracuse.

The model will be solved numerically using a standard numerical integration method (e.g., Runge-Kutta) to analyze the temporal evolution of $R(t), P(t)$ and $A(t)$. We also derived some analytical results. The results will be interpreted in the context of historical accounts to evaluate the plausibility of different scenarios. This approach allows us to quantify the relative importance of different siege dynamics and assess whether resource depletion, military pressure, or defensive capabilities were the dominant factors in the fall of Syracuse. In the next section we will analyze some scenarios concerning the siege of Syracuse, in order to explore the impact of various factors on the duration and outcome of the siege.


\section{Results}

\qquad Regarding the model defined in the previous section, Eq. \eqref{eq1} can be directly integrated to obtain the time evolution of the available resources inside the city:
\begin{equation} \label{eq4}
R(t) = \frac{\beta}{\alpha} + \left(R(0) - \frac{\beta}{\alpha}\right)e^{-\alpha\,t}
\end{equation}
\noindent
where $R(0)$ represents the initial amount of resources at $t=0$. This result indicates that $R(t)$ exponentially decays toward the stationary value  $\beta/\alpha$, which represents the long-term balance between resource depletion and resupply.

From Eq. \eqref{eq3}, we find two steady-state solutions: $A = 0$ and $P = \mu/\lambda$. Using the steady-state value of $R$, the long-time limit of Eq. \eqref{eq2} gives us:  
\begin{equation} \label{eq5}  
P = \frac{\delta\beta}{\alpha(\gamma+\kappa A)}.  
\end{equation}  

For the case where $A = 0$, we obtain $P = \frac{\delta\beta}{\alpha\gamma}$, leading to the first stationary solution:
\begin{equation} \label{eq6}  
(R, A, P) = \left(\frac{\beta}{\alpha}, 0, \frac{\delta\beta}{\alpha\gamma}\right).  
\end{equation}  

On the other hand, if $A \neq 0$, we found $P = \frac{\mu}{\lambda}$. However, this expression does not represent a physical stationary solution for $P$, as it contradicts physical expectations. In fact, for increasing values of $\mu$ (Roman army reinforcement rate), we would expect a decrease in the population of Syracuse $P$, not an increase as this expression suggests. Similarly, increasing $\lambda$ (the effectiveness of Syracuse's defenses) should enhance the survival of the population, implying an increase in $P$, not a decrease. The expression $P = \frac{\mu}{\lambda}$ merely identifies the condition under which the Roman army stabilizes (i.e., $dA/dt=0$), but it does not correspond to the actual steady-state value of $P$. The latter must be obtained from Eq. \eqref{eq5}, and it depends on the equilibrium value of $A$, as well as on some of the model's parameters.

However, the condition for the existence of a steady-state solution with $A\neq 0$ can be obtained through a stability analysis of the model equations \eqref{eq1} - \eqref{eq3} (see Appendix \footnote{The mathematical approach is grounded in standard dynamical systems theory \cite{holmes,strogatz,khalil}, and reveals a critical transition associated with a change in the stability of the Roman army's presence.}). This analysis reveals the presence of a critical defense threshold, $\lambda_c$, given by:
\begin{equation}  \label{eq9}  
\lambda_c = \frac{\alpha\gamma\mu}{\delta\beta}.  
\end{equation}

As shown in the Appendix, the stationary solution $A=0$ is stable only when $\lambda>\lambda_c$, i.e., when the effectiveness of Syracuse's defenses is sufficiently strong to eliminate the Roman army. On the other hand, for $\lambda<\lambda_c$, the Syracuse's  defenses are too weak to resist the Roman assault, allowing the invading force to persist and eventually prevail.

The derived critical value $\lambda_c$ represents a key threshold in the siege dynamics. $\lambda > \lambda_c$, Syracuse would have been able to resist and potentially defeat the Roman forces. On the other hand, if $\lambda < \lambda_c$, Roman forces would ultimately overpower the defenders, leading to the historical outcome of Syracuse's fall.

This threshold effectively determines whether the city's defenses, including fortifications and counterattacks (represented by $\lambda$), are strong enough to neutralize the advancing Roman army, given the siege's resource constraints and reinforcement dynamics.  The value of $\lambda_c$ is influenced by several key parameters, each corresponding to different aspects of the siege:

\begin{itemize}

\item $\alpha$ (rate of resource depletion): Higher values of $\alpha$ mean that Syracuse's supplies are exhausted more rapidly, making sustained defense more difficult. As $\alpha$ increases, $\lambda_c$ increases, meaning a stronger defense (higher $\lambda$) would be required to prevent defeat.

\item $\gamma$ (rate of population loss): Represents the rate at which the population of Syracuse declines due to starvation, disease, or casualties. A higher $\gamma$ makes it harder to maintain defense efforts, increasing $\lambda_c$ and making survival less likely.

\item $\mu$ (Roman reinforcement rate): Higher values of $\mu$ indicate that Rome can replenish its troops efficiently, sustaining its offensive pressure. If Rome had a strong logistical network providing constant reinforcements, $\lambda_c$ would rise, demanding an even stronger defense for Syracuse to survive.

\item $\delta$ (effectiveness of resources in supporting population): Determines how effectively resources ($R$) sustain the population ($P$). Higher $\delta$ means that fewer resources are needed to support the defenders, reducing $\lambda_c$, thus improving Syracuse's chances of survival.

\item $\beta$ (rate of resource resupply): Represents the external or internal resupply rate of Syracuse's resources. A higher $\beta$ allows Syracuse to sustain its population and defense for longer, reducing $\lambda_c$ and improving survival prospects.

\end{itemize}

And regarding the historical alignment? The actual historical scenario suggests that $\lambda < \lambda_c$, meaning Syracuse’s defenses, despite their ingenuity (including Archimedes’ war machines), were ultimately insufficient to counteract Roman military power, supply chains, and persistence. If Syracuse had access to greater supplies ($\beta$), suffered fewer losses ($\gamma$), or had a more effective defensive strategy ($\lambda$), its survival might have been possible. Conversely, Rome’s success was largely due to its ability to reinforce troops ($\mu$) and maintain pressure over the prolonged two-year siege.  

Summarizing, the value $\lambda_c$ serves as a decisive boundary separating victory from defeat. By analyzing how different factors influence $\lambda_c$, we gain deeper insight into the siege’s outcome and explore potential alternate historical scenarios. This result underscores the interplay between resource management, military strategy, and logistical capabilities in determining the fate of ancient cities under siege.

\begin{figure}[t]
\begin{center}
\vspace{6mm}
\includegraphics[width=0.48\textwidth,angle=0]{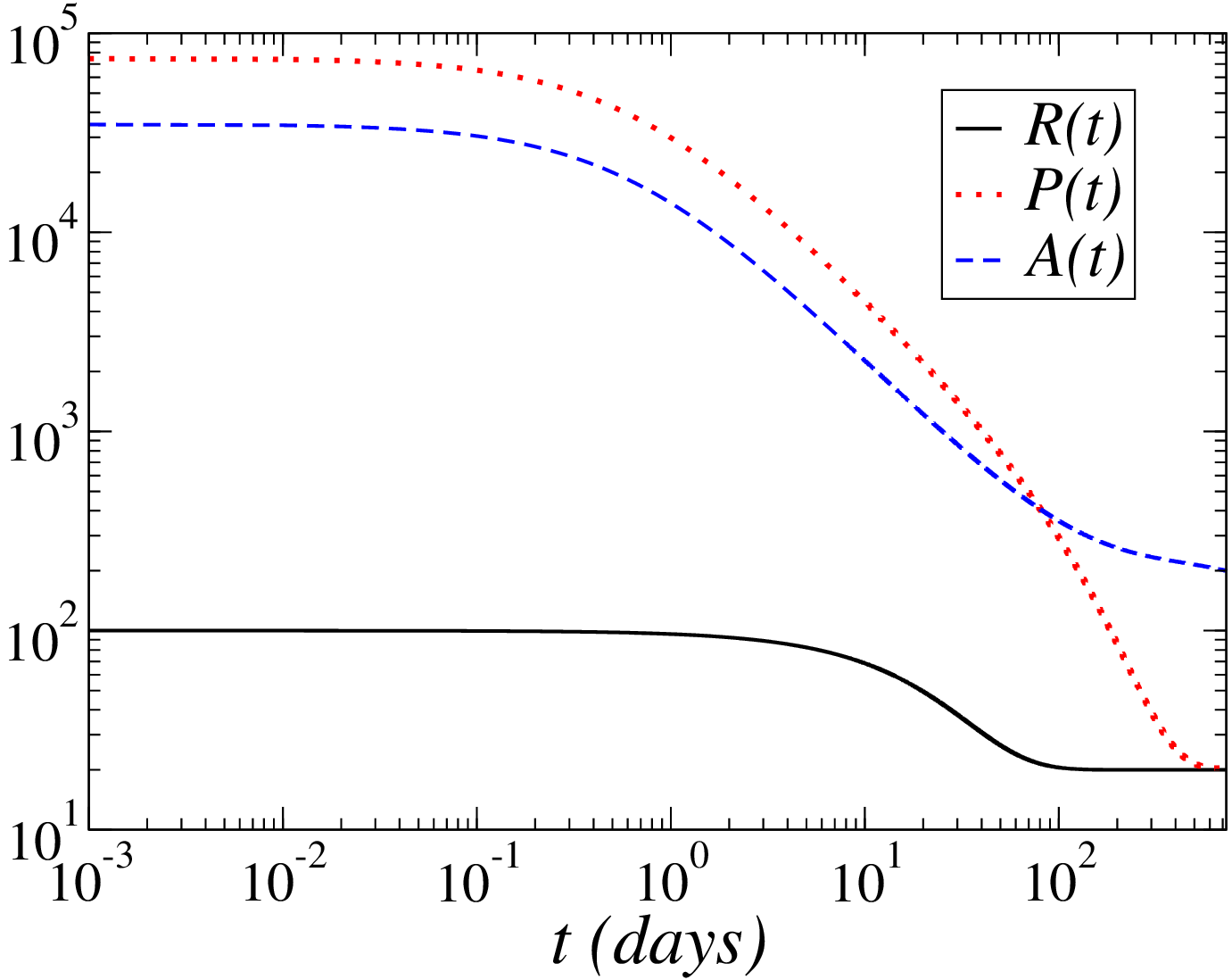}
\end{center}
\caption{Time evolution of the quantities $R(t), P(t)$ and $A(t)$. The time scale has been adjusted to match the duration of the Siege of Syracuse, approximately 2 years (730 days). The graph is presented in a log-log scale for better visualization. The parameters used are $\alpha=0.05, \beta=1.0, \gamma=0.001, \delta=0.01, \kappa=0.000043, \mu=0.0001$ and $\lambda=0.00002$.}
\label{fig1}
\end{figure}

Determining the initial conditions for our model presents a challenge due to the lack of precise historical data regarding the population of Syracuse and the size of the Roman army at the time of the siege. Ancient sources, such as Polybius \cite{Polibius} and Livy \cite{livy}, provide qualitative descriptions of the events but lack concrete figures. Plutarch's Live book \cite{plutarch} offers additional insights into the role of the Roman commander, but again, numerical data is scarce. Modern analyses, such as those by Goldsworthy \cite{goldsworthy_2001} and Lazenby \cite{Lazenby2}, attempt to reconstruct army sizes and logistical aspects of the war but often rely on estimations. Given these uncertainties, we consider reasonable approximations based on the available historical literature, allowing us to explore plausible scenarios for the siege. Thus, we considered as initial conditions $A(0)=35,000, P(0)=75,000$ and for $R(0)$ we choose a random numerical value, $R(0)=100.0$.

An illustration of the time evolution of the quantities $R(t), P(t)$ and $A(t)$ is shown in Fig. \ref{fig1}. The time scale has been adjusted to match the duration of the siege of Syracuse, approximately 2 years (730 days). The graph highlights the decline of Syracuse's population and the persistence of the Roman army. The depletion of resources due to the siege also plays a crucial role in the fall of Syracuse.

The proposed model provides insights into the key dynamics that influenced the Siege of Syracuse and allows for a quantitative analysis of its historical outcomes. Below, we discuss the implications of the model’s parameters and scenarios in the context of historical siege warfare.

\begin{enumerate}

\item The influence of resources on resistance duration: The depletion of resources ($\alpha$) and the possibility of resupply ($\beta$) play a crucial role in determining the length of the siege. A higher depletion rate with no external resupply leads to a faster collapse of the population, reinforcing historical accounts that emphasize the effectiveness of prolonged blockades in siege warfare. The presence of even minimal external aid ($\beta > 0$) could have prolonged resistance, suggesting that Syracuse’s isolation was a decisive factor in its downfall.

\item The balance between attack and defense: The parameter $\lambda$ encapsulates the effectiveness of Syracuse’s defenses, including the military innovations attributed to Archimedes. A high initial $\lambda$ would indicate that Rome faced significant resistance in direct assaults, which aligns with historical reports of Roman struggles against Syracusan countermeasures. However, as $\lambda$ potentially declined over time due to Roman tactical adaptations, the city’s ability to hold out diminished. This suggests that while superior defensive technology can delay a siege, it must be continuously adapted to remain effective.

\item The impact of Roman aggression: The term $-\kappa PA$ in the population equation shows that the level of Roman aggression directly influenced the rate of Syracusan casualties. If $\kappa$ was high, aggressive Roman assaults would have led to a faster population decline, potentially shortening the siege. If $\kappa$ was low, it implies that the Romans relied more on starvation and attrition rather than direct combat. Since the siege lasted two years, our model suggests that Rome may have initially adopted a more patient strategy before executing a final decisive attack.

\item Alternative outcomes and hypothetical scenarios: One of the key strengths of this model is its ability to explore counterfactual scenarios. If Carthage had successfully intervened and supplied Syracuse ($\beta > 0$), or if internal defenses had been sustained at a high level ($\lambda$ remained high), the city might have withstood the siege significantly longer. The model also allows us to test different values of $\kappa$ to assess whether a more aggressive Roman strategy could have shortened the siege further.

\end{enumerate}

Beyond the specific case of Syracuse, this model provides a framework for analyzing other historical sieges. The interplay between resource availability, military pressure, and defensive resilience is a common theme in siege warfare, and similar mathematical approaches could be used to study events such as the sieges of Carthage, Constantinople, or even modern urban warfare scenarios.

By formalizing these relationships in a mathematical model, we gain a structured way to compare different sieges and assess how variations in strategy and logistics influence their outcomes. This highlights the broader applicability of mathematical modeling in historical military analysis.

We can also explore some scenarios concerning the siege of Syracue. We can consider different scenarios to explore the impact of various factors on the duration and outcome of the siege, namely:

\begin{itemize}

\item \textit{Complete blockade vs. limited resupply:} $\beta = 0$ models a perfect blockade, while $\beta > 0$ represents some level of external support.

\item \textit{Impact of starvation and resource scarcity:} A high $\alpha$ (rapid depletion of resources) combined with a high $\gamma$ (population decline due to starvation) may lead to a faster collapse.

\item \textit{Roman siege strategy:} A high $\kappa$ represents aggressive Roman attacks, whereas a low $\kappa$ models a more patient, attritional siege approach.

\item \textit{Potential external intervention:} If Carthage had successfully provided reinforcements, it could be modeled by adjusting $\beta$ (resupply rate) or introducing an additional term to reflect counterattacks.

\end{itemize}

\begin{figure}[t]
\begin{center}
\vspace{6mm}
\includegraphics[width=0.4\textwidth,angle=0]{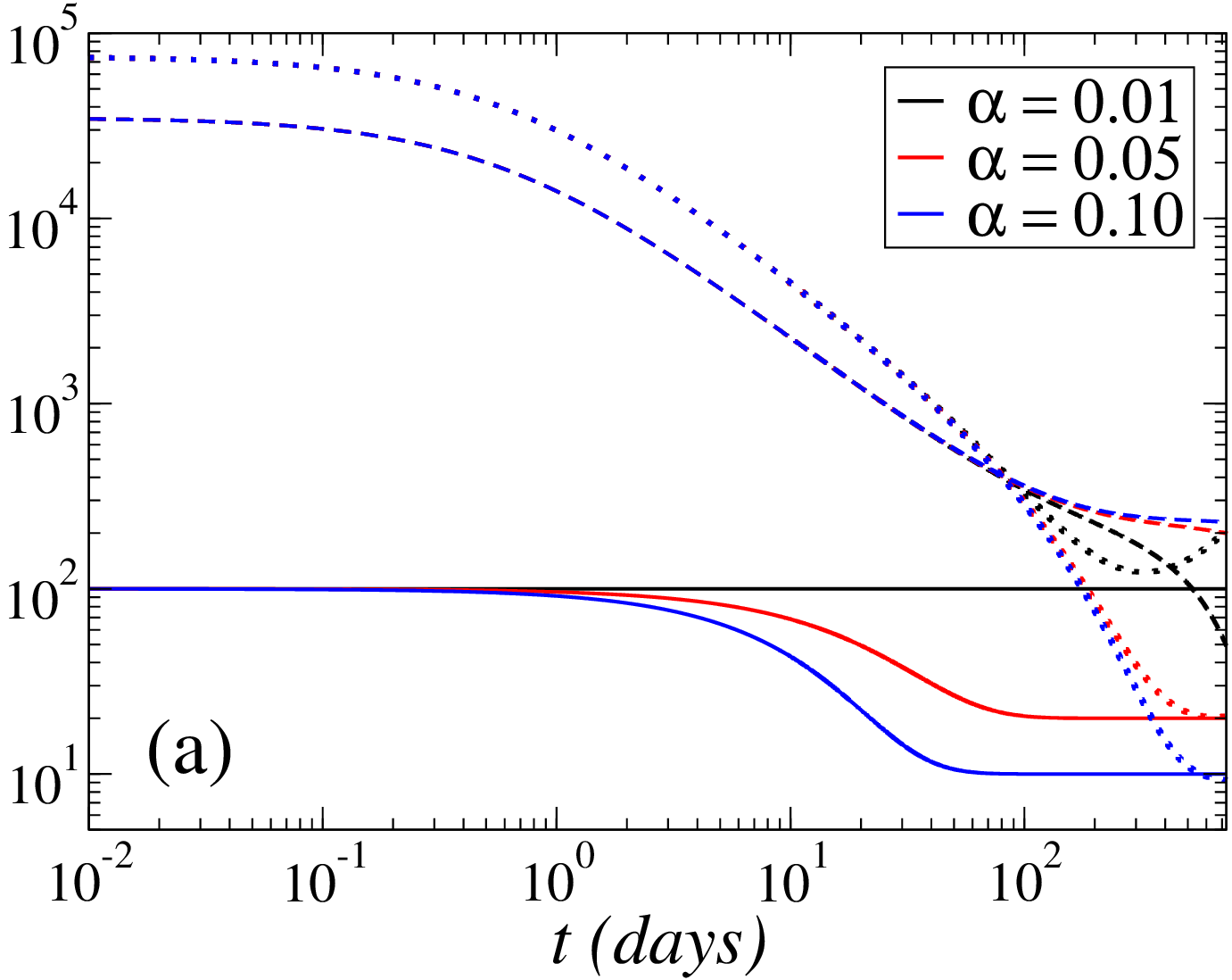}
\hspace{0.3cm}
\includegraphics[width=0.4\textwidth,angle=0]{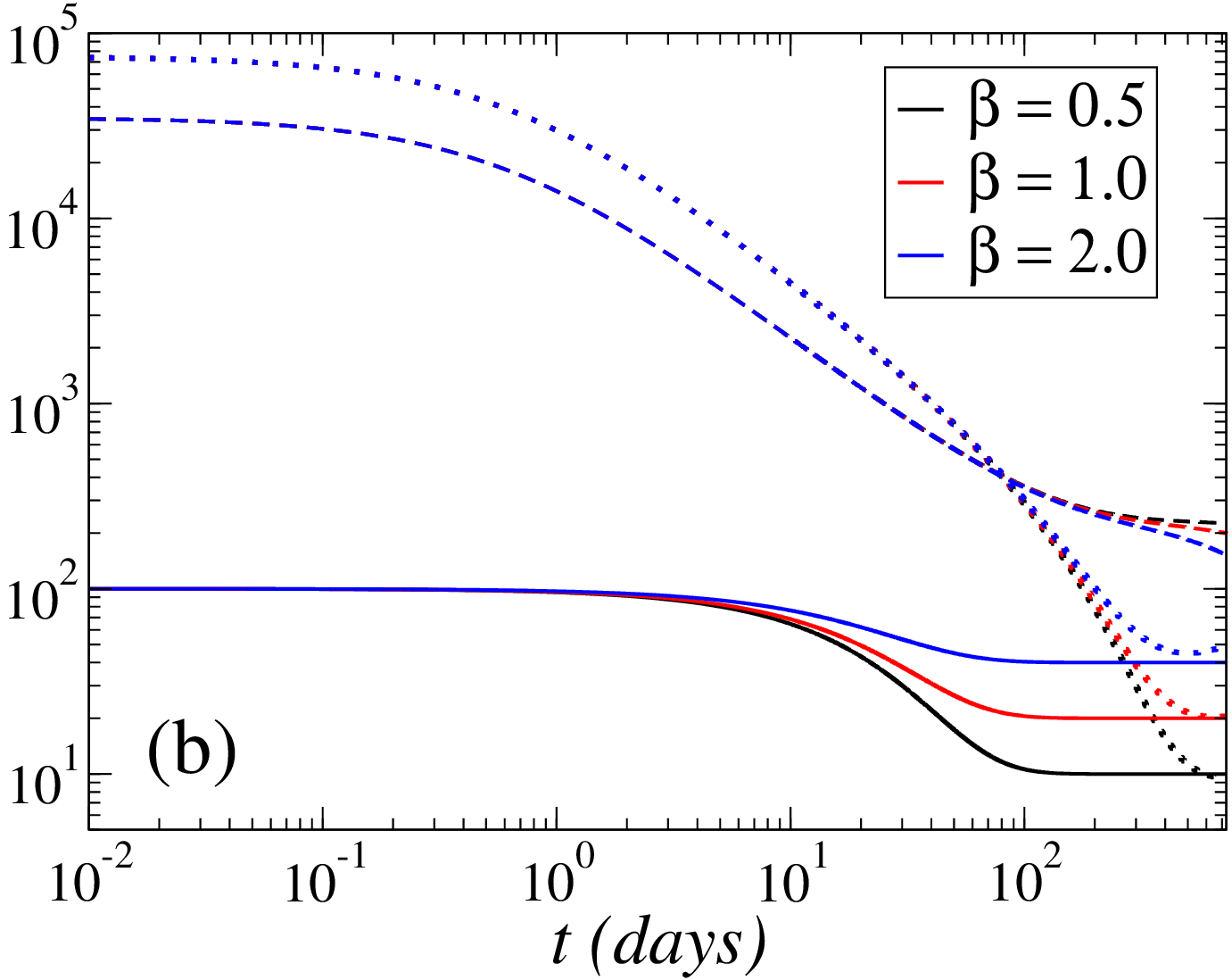}
\\
\vspace{1.0cm}
\includegraphics[width=0.4\textwidth,angle=0]{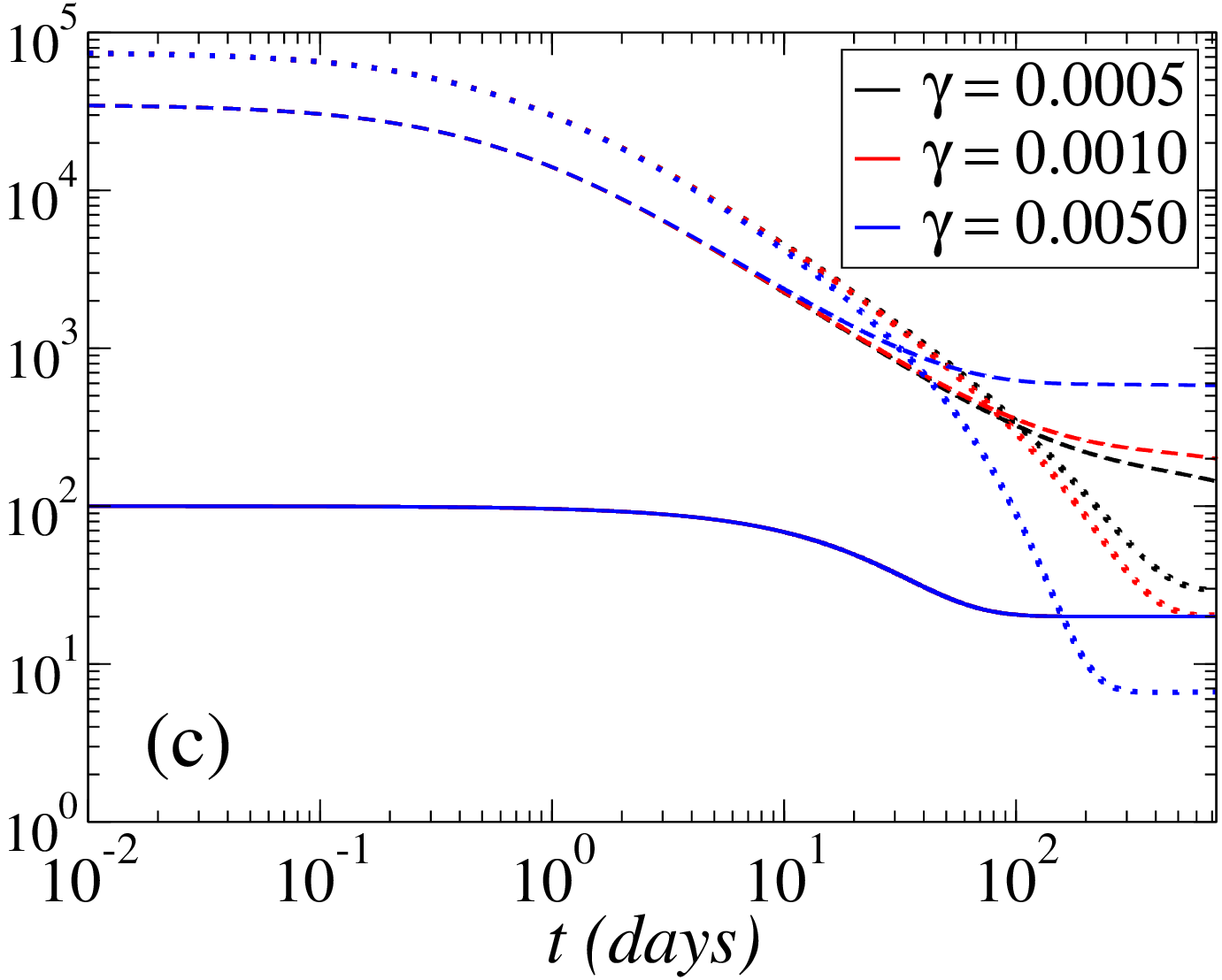}
\hspace{0.3cm}
\includegraphics[width=0.4\textwidth,angle=0]{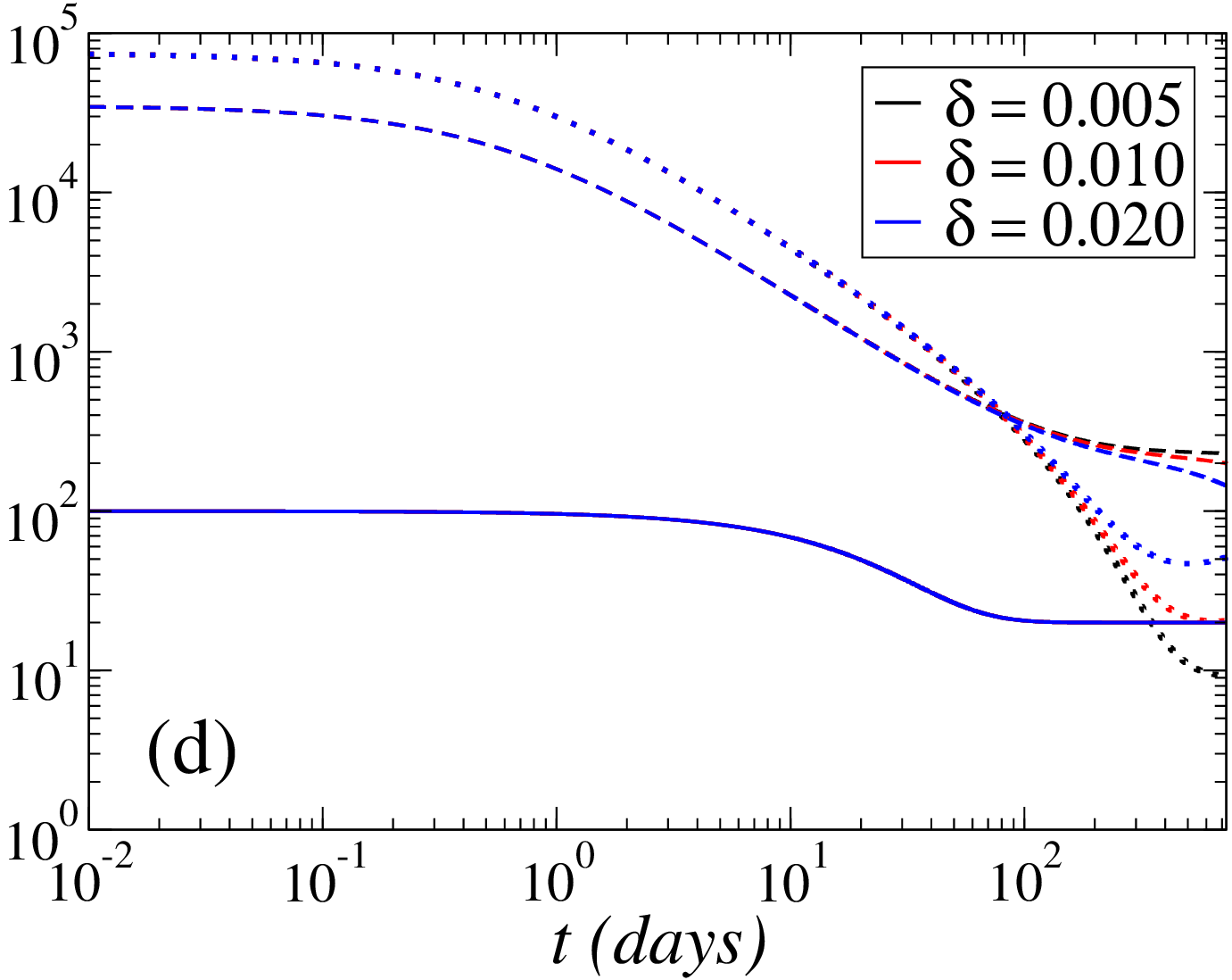}
\end{center}
\caption{Time evolution of the quantities $R(t), P(t)$ and $A(t)$. The full lines represent $R(t)$, the dotted lines stand for $P(t)$ and the dashed ones represent $A(t)$. We considered the same parameters as in Fig. \ref{fig1}, but we varied separately one of them to analyse scenarios. The varied parameters in each figure are: (a) $\alpha$; (b) $\beta$; (c) $\gamma$; (d) $\delta$.}
\label{fig2}
\end{figure}

\begin{figure}[t]
\begin{center}
\vspace{6mm}
\includegraphics[width=0.4\textwidth,angle=0]{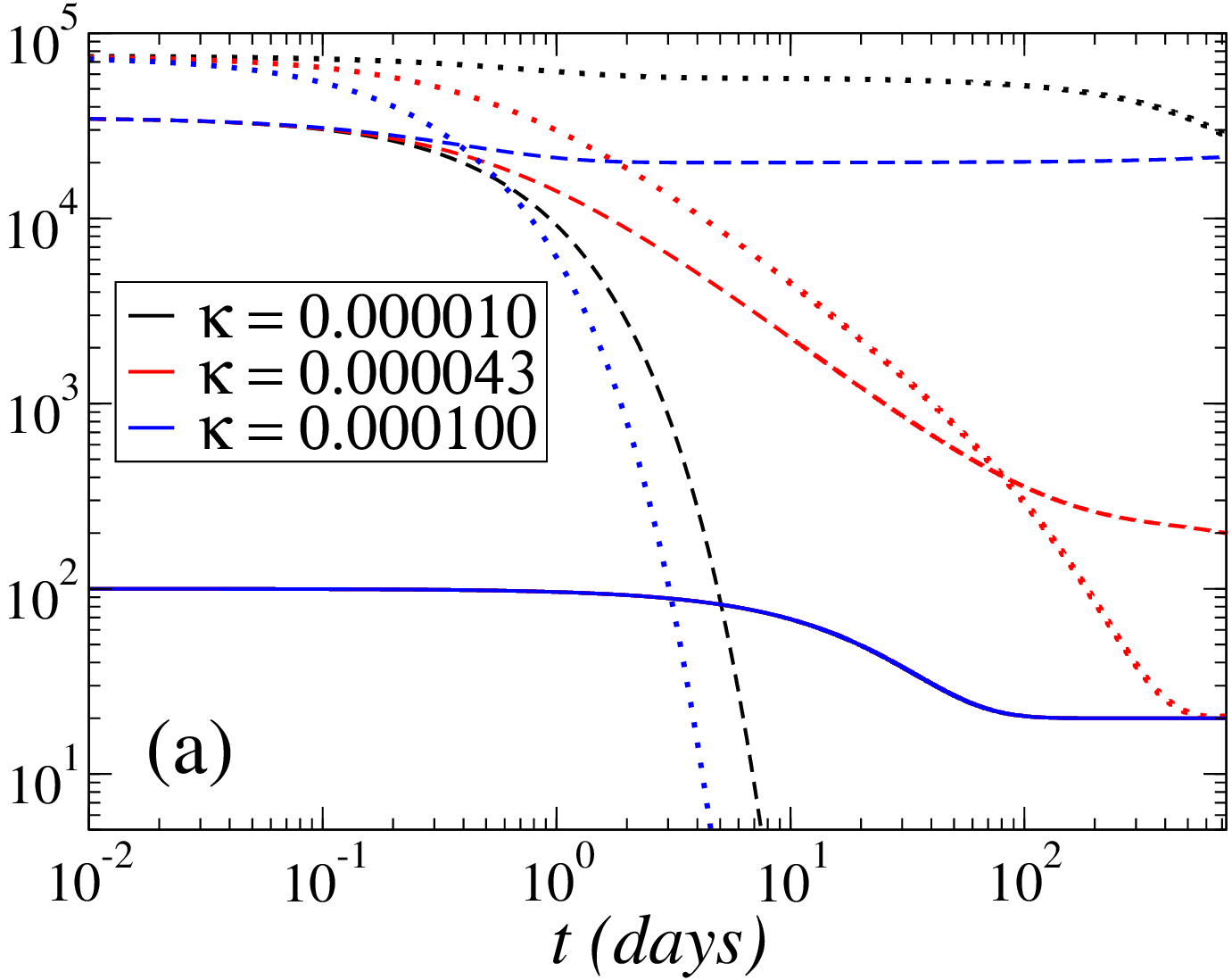}
\hspace{0.3cm}
\includegraphics[width=0.4\textwidth,angle=0]{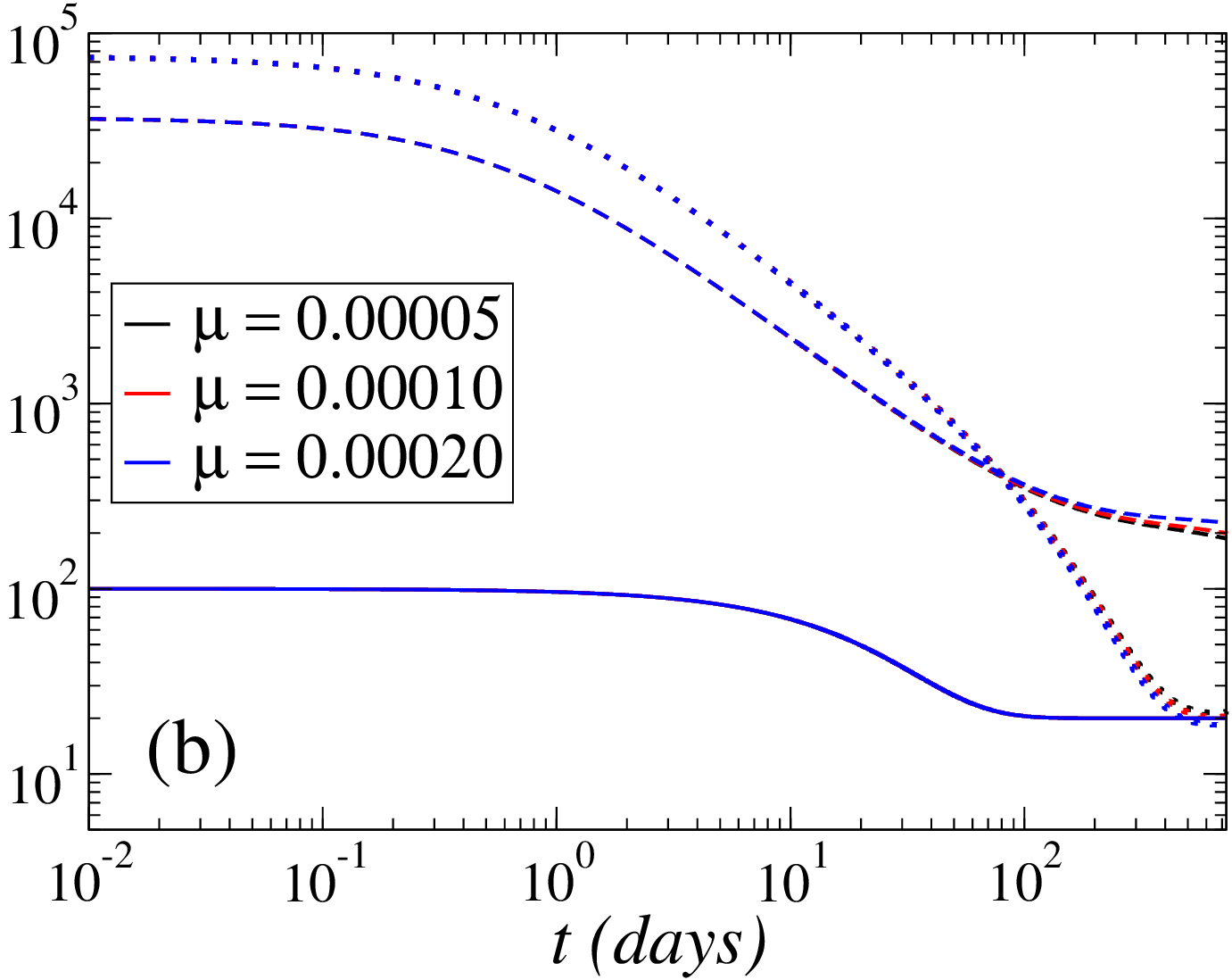}
\\
\vspace{1.0cm}
\includegraphics[width=0.4\textwidth,angle=0]{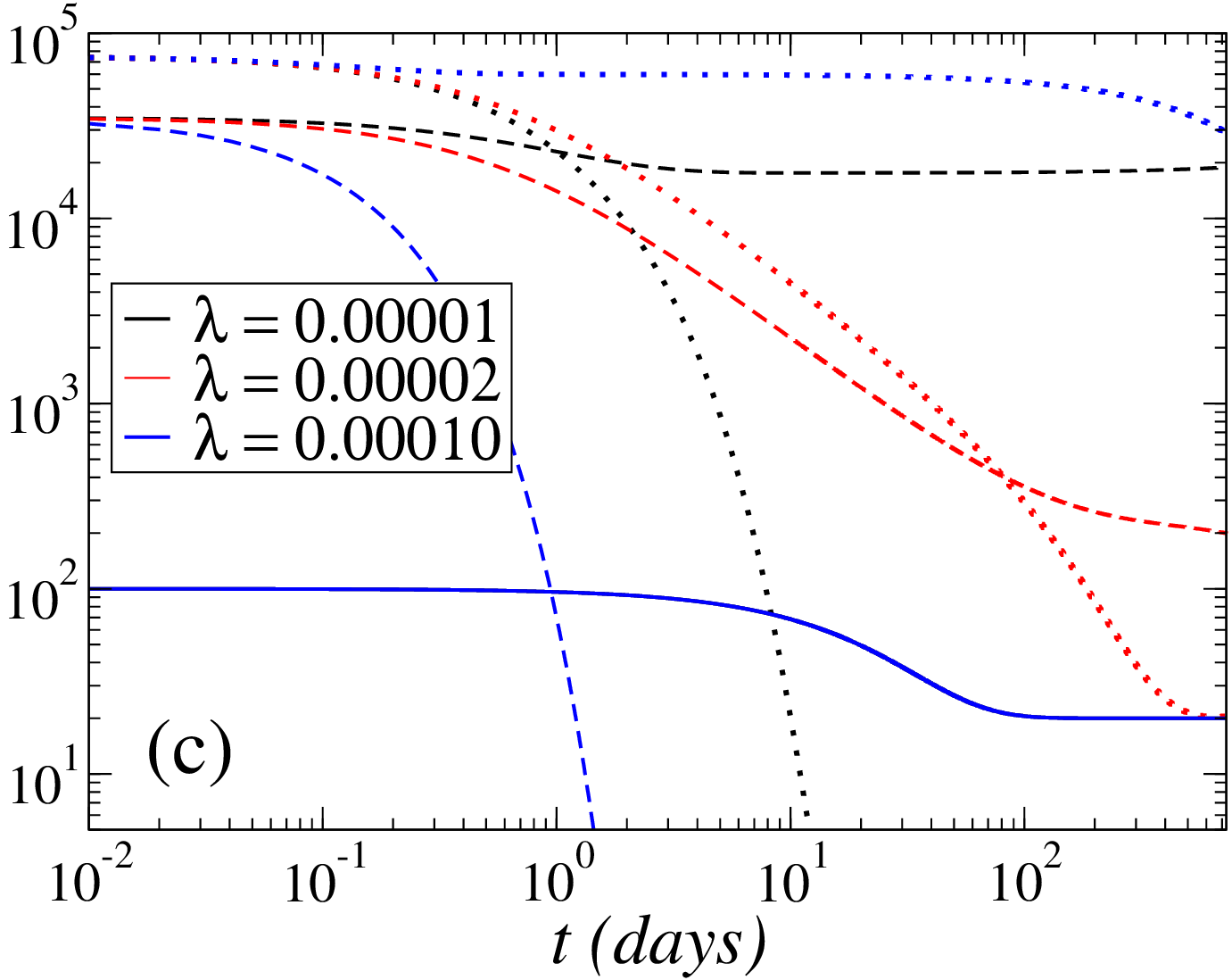}
\end{center}
\caption{Time evolution of the quantities $R(t), P(t)$ and $A(t)$. The full lines represent $R(t)$, the dotted lines stand for $P(t)$ and the dashed ones represent $A(t)$. We considered the same parameters as in Fig. \ref{fig1}, but we varied separately one of them to analyse scenarios. The varied parameters in each figure are: (a) $\kappa$; (b) $\mu$; (c) $\lambda$.}
\label{fig3}
\end{figure}

Let's explore some ``what-if'' scenarios regarding the dynamics of the siege. Figures \ref{fig2} and \ref{fig3} present numerical results obtained by varying a single parameter at a time while keeping all others fixed, allowing us to isolate and analyze the specific impact of each parameter on the system's behavior. First, in Fig. \ref{fig2}, panel (a), we varied $\alpha$, that represents the rate at which available resources within the city are depleted due to the siege. It directly influences the duration for which Syracuse can sustain itself under Roman pressure. For $\alpha = 0.01$ (slow resource depletion), resources remain available for a longer period. This scenario suggests that Syracuse could have resisted the siege for a significantly extended time, potentially delaying its fall. The population decline is also slower, as resource availability allows for a more prolonged survival of the inhabitants. However, if the Roman army is persistent enough, the city might still fall, but only after a much longer conflict. For $\alpha = 0.05$ (moderate resource depletion - reference case), we have the baseline case where the siege duration is roughly aligned with historical records ($\sim 2$ years). Resources decline at a reasonable rate, and the fall of Syracuse happens within a timeframe consistent with historical estimates. The balance between resource loss and population decline suggests that the model captures the key dynamics of the siege accurately. Finally, for $\alpha = 0.10$ (fast resource depletion), resources are exhausted much more quickly, accelerating the collapse of Syracuse. The population declines faster due to starvation and worsening conditions. In this scenario, the city would likely have fallen significantly earlier than in historical records, indicating that such a high value for $\alpha$ may not be realistic given the known duration of the siege. Thus, the faster the resource depletion ($\alpha$ high), the sooner Syracuse falls.  If $\alpha$ is too small, the city can resist for a much longer time, making the Roman victory uncertain. The \textit{historical duration} of $\sim 2$ years suggests an intermediate value, supporting the plausibility of the reference choice ($\alpha = 0.05$).

In Fig. \ref{fig2}, panel (b), we varied $\beta$, that accounts for the influx or generation of resources within Syracuse during the siege. This could be due to stockpiled supplies, local production, or limited external aid. A higher $\beta$ means that resources are replenished at a greater rate, potentially prolonging resistance. For $\beta = 0.5$ (low resource influx), resources are replenished at a very low rate, insufficient to sustain the population and defenses for long. As a result, resource depletion ($R(t)$) happens more rapidly, leading to an accelerated collapse of the city. The population declines quickly due to famine and the inability to maintain internal stability. The city would fall faster than the historical estimate of two years. For $\beta = 1.0$ (moderate resource influx - reference case), we have the baseline case, leading to a siege duration close to the historical record. The balance between resource loss and replenishment allows the city to survive for about two years before succumbing. This suggests that Syracuse had some ability to manage supplies but was ultimately unable to outlast the Romans. Finally, for $\beta = 2.0$ (high resource influx), the city can sustain itself for a much longer time. In this case, Syracuse’s resistance is significantly extended, possibly altering the historical outcome. The population decline is much slower, as resources are more available. If the defenses ($\lambda$) were also sufficiently strong, this scenario could have led to a successful defense against the Romans. Thus, a low $\beta$ results in a quick defeat, suggesting that resource replenishment was crucial for prolonging resistance. A moderate $\beta$ matches the \textit{historical timeline}, supporting the idea that Syracuse had limited but not entirely negligible resources. A high $\beta$ would allow Syracuse to resist for much longer, making a Roman victory uncertain or at least significantly delayed.

In Fig. \ref{fig2}, panel (c), we varied $\gamma$, that controls the rate at which the population of Syracuse declines over time due to internal and external pressures. A higher $\gamma$ means that the population diminishes more rapidly, reducing the city's ability to sustain its defenses and workforce. For $\gamma = 0.0005$ (slow population decline), the city's population remains relatively stable for a longer period, maintaining its defensive capabilities. This results in a more prolonged resistance, as there are still enough people to sustain essential functions inside the city. However, if resources ($R$) eventually run out, collapse will still occur, but more gradually.  For $\gamma = 0.0010$ (moderate population decline - reference case), we have the baseline scenario, leading to a population decline over approximately two years. This balance reflects the historical case where the city resisted for a significant duration but ultimately could not sustain itself. Finally, for $\gamma = 0.0050$ (fast population decline), the population of Syracuse diminishes much faster, accelerating the fall of the city. A rapid loss of people weakens internal organization and reduces the effectiveness of defensive strategies ($\lambda$). Even if resources ($\beta$) are adequate, a high mortality rate makes prolonged resistance unfeasible. Thus, a lower $\gamma$ allows the city to sustain its population for a longer time, delaying the fall. A moderate $\gamma$ leads to a scenario compatible with \textit{historical records}. A high $\gamma$ results in a quicker collapse, suggesting that an increased impact of starvation, disease, or other internal pressures could have shortened the siege significantly.

In Fig. \ref{fig2}, panel (d), we varied $\delta$, that governs how effectively the available resources support the population of Syracuse. Higher values indicate that resources contribute more efficiently to sustaining the population, whereas lower values mean that even with sufficient resources, the population might still decline due to inefficiencies, distribution issues, or internal conflicts. For $\delta = 0.0050$ (low resource effectiveness), even if resources ($R$) are available, they do not significantly help sustain the population ($P$). This could represent logistical problems, corruption, or spoilage of food supplies. The population declines faster, increasing the likelihood of an early collapse of Syracuse. For $\delta = 0.010$ (moderate resource effectiveness - reference case), we have the baseline scenario, where resources effectively contribute to population maintenance, but only to a limited extent. The population decline is slowed, but if resources deplete reaches a critical level, Syracuse still falls within a realistic historical timeframe. For $\delta = 0.020$ (high resource effectiveness), resources strongly sustain the population. As long as some resources remain, the population persists for a longer time. This scenario suggests that, had Syracuse been able to use its resources more efficiently (e.g., through better rationing, supply chains, or external aid), the siege could have lasted much longer. Thus, a lower $\delta$ means that even with available resources, the population struggles to survive, leading to an earlier fall. A moderate $\delta$ aligns with \textit{historical expectations}, allowing resistance for a reasonable period before collapse. A higher $\delta$ suggests that with better logistics or supply management, Syracuse might have resisted for much longer, potentially changing the course of the siege.

In Fig. \ref{fig3}, panel (a), we varied $\kappa$, that models the lethality or pressure exerted by the Roman forces on the inhabitants of Syracuse. A higher value implies that direct conflict, bombardment, or other military actions quickly diminish the population. A lower value suggests that the Roman forces were less efficient in directly harming the population, either due to tactical choices, defensive measures, or logistical constraints. For $\kappa = 0.00001$ (low military pressure), the Roman attacks are relatively ineffective at reducing the population. The main threats to Syracuse’s survival come from resource depletion ($\alpha$) and other internal factors rather than direct military losses. This could correspond to an early phase of the siege, when the Romans faced difficulties in breaching the defenses or conducting effective assaults. For $\kappa = 0.000043$ (baseline scenario), we have a reasonable effectiveness of Roman attacks, leading to population decline over time. While resource depletion and other factors contribute to the collapse, the direct impact of military actions is non-negligible. This aligns with the known \textit{historical accounts}, where the siege lasted for two years, but over time, increasing pressure from the Romans weakened the defenders. For $\kappa = 0.0001$ (high military pressure), the Roman attacks cause rapid population decline. Even if resources remain available, the direct effect of military aggression leads to the fall of Syracuse in a much shorter timeframe. This scenario could represent a hypothetical case in which the Romans employed more aggressive siege tactics, such as earlier breaches, continuous assaults, or overwhelming force, leading to a quick collapse of the city's resistance. Thus, a low $\kappa$ suggests that factors other than military confrontation, such as starvation or disease, played a more significant role in the fall of Syracuse. A moderate $\kappa$ aligns with the \textit{historical timeframe}, where the Roman attacks contributed to the eventual collapse but did not immediately overwhelm the defenders. A high $\kappa$ suggests a much shorter siege, indicating that if the Romans had been more aggressive or effective in their assaults, the city could have fallen significantly earlier.  

In Fig. \ref{fig3}, panel (b), we varied $\mu$, that directly influences the resilience of the Roman army ($A$). A higher value means that the Romans are capable of sustaining their military presence despite losses, while a lower value suggests difficulties in maintaining or replacing troops over time. For $\mu = 0.00005$ (low reinforcement rate), the Roman army struggles to replenish itself. If Syracuse’s defenses ($\lambda$) are strong enough, there is a possibility that the Romans could be repelled over time. This case corresponds to a scenario where logistical difficulties, disease, desertions, or losses in battle significantly weaken the Roman presence. For $\mu = 0.0001$ (baseline scenario), the Roman army is replenished at a moderate rate, which aligns with historical reality. The army does not grow indefinitely, but it also does not vanish due to attrition. This case represents a \textit{prolonged} but ultimately \textit{successful} siege, where the Romans maintain a persistent presence despite setbacks. For $\mu = 0.0002$ (high reinforcement rate), the Roman army is quickly replenished, making it nearly impossible for Syracuse’s defenders to deplete the attacking forces. Even if the city had more resources and stronger defenses, the overwhelming ability of Rome to bring in new troops ensures that the siege will succeed in a much shorter timeframe. This scenario would align with an alternate history where the Romans committed far more resources to the siege, leading to a swift and decisive conquest. Thus, a low $\mu$ suggests that if Syracuse had held out long enough, Rome might have abandoned the siege due to unsustainable losses. A moderate $\mu$ aligns with the \textit{historical outcome}, where the Roman army was persistent but not invulnerable, leading to a \textit{gradual conquest}. A high $\mu$ suggests that the fall of Syracuse was inevitable, as Rome's ability to maintain a constant army would ensure eventual success, regardless of other factors.

In Fig. \ref{fig3}, panel (c), we varied $\lambda$, that quantifies how effectively Syracuse’s defenses can neutralize the Roman forces. A higher $\lambda$ means that the defenders can inflict significant damage on the attacking army, potentially prolonging the siege or even repelling the Romans if the defenses are strong enough. For $\lambda = 0.00001$ (weak defenses), the defenses of Syracuse are largely ineffective. The Roman army ($A$) suffers minimal losses, leading to a rapid conquest of the city. This could correspond to a case where the defensive structures were compromised, the morale of the defenders was low, or the innovations of Archimedes were not effectively deployed. For $\lambda = 0.00002$ (baseline scenario), it represents a scenario in which the defenses are moderately effective, slowing the Roman advance but not stopping it completely. This case aligns with \textit{historical accounts}, where Archimedes' war machines and other fortifications initially caused significant Roman losses but ultimately failed to prevent the city's fall. The siege is prolonged, but Rome eventually prevails. For $\lambda = 0.00010$ (strong defenses), Syracuse's defenses are highly effective, dealing substantial damage to the Roman army. If $\lambda$ surpasses the critical threshold $\lambda_c$, the Roman forces could potentially be defeated, leading to a scenario in which the siege is broken and the city survives. This would align with an alternate history where Archimedes' inventions played a decisive role, perhaps combined with stronger walls, better coordination, or reinforcements from allies. Thus, a low $\lambda$ makes the fall of Syracuse inevitable, with minimal resistance to Roman advances. A moderate $\lambda$ matches the historical outcome, where Syracuse put up a significant fight but ultimately succumbed. A high $\lambda$ suggests an alternate history where the defenders could have successfully repelled the Romans, especially if other factors (e.g., resource availability, reinforcements) were also in their favor.


\section{Final Remarks}   

\qquad In this work we considered a mathametical model for analyzing the siege of Syracuse. For this purpose, we propose a set of three ordinary differential equations for the time evolution of three key quantities, namely the amount of available resources inside Syracuse $R(t)$, the population of Syracuse $P(t)$, including both civilians and defenders, and the size of the attacking Roman army $A(t)$.

The fall of Syracuse was not an inevitable outcome from the outset of the siege. Instead, it was the result of a complex interplay of multiple factors, including resource depletion, the effectiveness of defensive strategies, Roman persistence, and the impact of external support-or lack thereof. The mathematical model presented in this study provides a framework to explore alternative historical scenarios and assess how changes in key parameters might have influenced the fate of the city.

By adjusting the values of parameters such as the rate of resource consumption ($\alpha$), the effectiveness of the Syracusan defenses ($\lambda$), the intensity of Roman attacks ($\kappa$) and the Roman's ability to reinforce troops ($\mu$), we can analyze plausible outcomes where Syracuse might have prolonged its resistance or, conversely, succumbed more rapidly. A scenario with greater external support ($\beta$ greater), for instance, could have extended the duration of the siege, potentially allowing for reinforcements from Carthage or other allies. Conversely, a higher depletion rate of resources ($\alpha$) or increased Roman aggression ($\kappa$) could have led to a quicker collapse.

These insights highlight the value of mathematical modeling in historical analysis. Rather than viewing past events as preordained, this approach allows for a systematic investigation of counterfactual scenarios, providing a deeper understanding of the dynamics that shaped historical conflicts. While the actual events that transpired are known, modeling offers a tool to dissect the intricate dependencies and explore how history might have unfolded under slightly different circumstances.


About the model's scope and limitation, the proposed mathematical model captures several key aspects of the siege of Syracuse. Below, we discuss which historical factors are directly incorporated into the model and which aspects are left implicit or omitted. About the elements explicitly represented in the model, one can cite:

\begin{itemize}

\item \textit{Resource depletion and limited resupply:} Indeed, the term $-\alpha\,R$ represents the natural depletion of resources within the besieged city due to consumption. The parameter $\beta$ accounts for potential external resupply, which can be interpreted as any support Syracuse might have received. If \(\beta\) is set to zero or a small value, this reflects an effective Roman blockade preventing reinforcements.

\item \textit{Decline in population due to starvation and attacks:} It can be viewed in the term $-\gamma\,P$ in the equation for $P(t)$, and represents population decline due to hunger and disease, common in prolonged sieges. Additionally, the term $+\delta\, R$ models the effect of resource availability on sustaining the population, assuming that better access to supplies slows the rate of decline.

\item \textit{Roman offensive capabilities}: The term $-\kappa PA$ explicitly represents the losses in population due to direct Roman attacks. This interaction ensures that the aggression of the Roman forces ($A$) directly influences the decline in defenders and civilians.

\item \textit{Roman army reinforcement: }The parameter $\mu$ represents the rate at which the Roman army is replenished. This replenishment could come from multiple sources, including reinforcements arriving by sea, recruitment of local mercenaries, or supply lines from Roman-controlled territories. Unlike the natural decay of armies due to attrition or battlefield losses,  $+\mu A$ introduces an external source of troops, ensuring that the Roman force does not diminish indefinitely. The magnitude of  depends on logistical factors, available reserves, and strategic decisions made by the Roman command.

\item \textit{The role of defensive measures, including Archimedes' inventions:} The term $-\lambda\,A\,P$ models the effect of Syracuse's defenses on Roman forces. Instead of introducing a separate parameter for technological advantages, we interpret $\lambda$ as an aggregate measure of all defensive efforts, including fortifications, strategic positioning, and Archimedes' war machines.

\end{itemize}

Particularly, the inclusion of the term $+\mu A$ in the equation for the attacking Roman forces introduces a crucial aspect of historical realism into the model. This term accounts for the replenishment of Roman troops during the siege, a well-documented feature of Roman military campaigns \cite{livy}. Without it, the model would assume a continuous depletion of the Roman forces, which does not accurately reflect their logistical capabilities and strategic persistence. By incorporating such term, the model recognizes that Rome had the resources to sustain its military presence for extended periods. This factor is consistent with historical accounts that describe Rome's ability to recruit and deploy fresh troops when necessary, preventing significant attrition from weakening their offensive. The replenishment of forces creates a more asymmetrical struggle, where Syracuse’s defenses (modeled through $\lambda$) must not only inflict losses but do so at a rate that exceeds the reinforcements. If $\lambda<\lambda_c$, the critical defense threshold, then the Roman forces grow over time, making the eventual fall of the city more likely.

In summary, our model captures the essential mechanisms of siege warfare while leaving out certain historical nuances that, while significant, would require a more detailed parameterization. The addition of the reinforcement term makes the model more historically faithful by incorporating the reality of Rome’s military logistics. While increasing complexity slightly, this refinement enhances the explanatory power of the model and offers a more realistic representation of the factors that led to Syracuse’s downfall.


\section*{Acknowledgments}

The author acknowledges partial financial support from the Brazilian scientific funding agency Conselho Nacional de Desenvolvimento Cient\'ifico e Tecnol\'ogico (CNPq, Grant 308643/2023-2). I am grateful for the fruitful discussions with the History teacher Prof. Juliana Sant'Anna.


\appendix
\section{Stability analysis}

The local stability of the model's stationary solution can be inferred from the eigenvalues of the Jacobian matrix $J$ obtained from the system of differential equations \eqref{eq1} - \eqref{eq3}. An equilibrium point is locally asymptotically stable if all eigenvalues of $J$ have negative real parts \cite{holmes,strogatz,khalil}. The eigenvalues can be obtained from $det(J-\Lambda\,I)=0$, where $I$ is the identity matrix. The Jacobian matrix is given by taking the partial derivatives of Eqs. \eqref{eq1} - \eqref{eq3}, namely

\begin{equation*}
J = 
\begin{bmatrix}
\frac{\partial{\dot{R}}}{\partial{R}} & \frac{\partial{\dot{R}}}{\partial{P}} & \frac{\partial{\dot{R}}}{\partial{A}} \\
\frac{\partial{\dot{P}}}{\partial{R}} & \frac{\partial{\dot{P}}}{\partial{P}} & \frac{\partial{\dot{P}}}{\partial{A}} \\
\frac{\partial{\dot{A}}}{\partial{R}} & \frac{\partial{\dot{A}}}{\partial{P}} & \frac{\partial{\dot{A}}}{\partial{A}}
\end{bmatrix}
\end{equation*}

Taking the derivatives, the matrix becomes

\begin{equation*}
J = 
\begin{bmatrix}
-\alpha & 0 & 0 \\
\delta & -\gamma-\kappa\,A & -\kappa\,P \\
0 & -\lambda\,A & \mu-\lambda\,P
\end{bmatrix}
\end{equation*}

For the solution $(R,A,P)=(\beta/\alpha,0,\delta\beta/\alpha\gamma)$, the eigenvelues of the Jacobian matrix are $\Lambda_{1}=-\alpha, \Lambda_{2}=-\gamma$ and $\lambda_{3}=\mu-\lambda\delta\beta/\alpha\delta$. We have $\Lambda_{1}<0$ and $\Lambda_{2}<0$ since $\alpha>0$ and $\gamma>0$. We have $\Lambda_3<0$ only if
\begin{equation}\label{app1}
\lambda>\frac{\alpha\gamma\mu}{\delta\beta}
\end{equation}
From Eq. \eqref{app1}, we identify
\begin{equation}\label{app2}
\lambda_c = \frac{\alpha\gamma\mu}{\delta\beta}
\end{equation}
\noindent
In other words, the solution $A=0$ (extinction of the Roman army) is stable only if $\lambda>\lambda_c$, which is the condition for the survival of Syracuse. Conversely, if $\lambda<\lambda_c$, the Roman army persists, ultimately leading to Syracuse's fall (which is related to a solution $A\neq 0$).

\bibliographystyle{elsarticle-num-names}

\begin{thebibliography}{00}



\bibitem{livy}
Livy. (c. 10 AD), \textit{History of Rome}, Penguin Classics.


\bibitem{plutarch}  
Plutarch, Lives, Vol. V: \textit{Agesilaus and Pompey. Pelopidas and Marcellus}, translated by B. Perrin (Harvard University Press, Cambridge, MA, 1917).



\bibitem{goldsworthy_2001}
A. Goldsworthy, \textit{The Punic Wars} (Cassell, London, 2001).


  

\bibitem{indi_movie}
J. Mangold (Director), \textit{Indiana Jones and the Dial of Destiny [Film]},  Lucasfilm Ltd., Walt Disney Studios Motion Pictures (2023).




\bibitem{Polibius}
Polybius, \textit{The Histories}, Volume I: Books 1–2 and Volume II: Books 3–4, , trans. by W. R. Paton,  Loeb Classical Library (Harvard University Press, Cambridge, MA, 1922).



\bibitem{Lazenby}
J. F. Lazenby, \textit{The First Punic War: A Military History} (Stanford University Press, 1996).



\bibitem{galam_book}
S. Galam, \textit{Sociophysics: A Physicist's Modeling of Psycho-political Phenomena} (Springer, Berlin, 2012).

  
\bibitem{csf}
M. G. E. da Luz, C. Anteneodo, N. Crokidakis, M. Perc, \textit{Sociophysics: Social collective behavior from the physics point of view}, Chaos, Solitons $\&$ Fractals 170, 113379 (2023).



\bibitem{sooknanan}
J. Sooknanan, D. M. G. Comissiong, \textit{When behaviour turns contagious: the use of deterministic epidemiological models in modeling social contagion phenomena}, Int. J. Dynam. Control 5, 1046-1050 (2017).



\bibitem{lanchester1}
F. W. Lanchester, \textit{Aircraft in Warfare: The Dawn of the Fourth Arm} (London: Constable and Company Ltd., 1916).


\bibitem{lanchester2}
F. W. Lanchester, \textit{Mathematics in Warfare}, in The World of Mathematics, Vol. 4, Ed. Newman, J. R., Simon and Schuster, pp. 2138–2157, 1956.


\bibitem{Richardson1}
L. F. Richardson, \textit{Arms and Insecurity: A Mathematical Study of the Causes and Origins of War} (Pittsburgh, Boxwood Press, 1960).



\bibitem{Richardson2}  
L. F. Richardson, \textit{Statistics of Deadly Quarrels} (Pittsburgh, Boxwood Press, 1960).



\bibitem{turchin1}
P. Turchin, \textit{Historical Dynamics: Why States Rise and Fall} (Princeton University Press, Princeton, 2003)



\bibitem{turchin2}
P. Turchin, \textit{War and Peace and War: The Life Cycles of Imperial Nations} (Pi Press, New York, 2005).



\bibitem{gunduz_empire}
G. Gunduz, \textit{The dynamics of the rise and fall of empires}, Int. J. Mod. Phys. C 27, 1650123 (2016).






\bibitem{holmes}
J. Guckenheimer, P. Holmes, \textit{Nonlinear Oscillations, Dynamical Systems, and Bifurcations of Vector Fields} (Springer, New York, 2002).


\bibitem{strogatz}
S. H. Strogatz, \textit{Nonlinear Dynamics and Chaos: With Applications to Physics, Biology, Chemistry, and Engineering}, 2nd ed. (CRC Press, 2015).

\bibitem{khalil}
H. K. Khalil, \textit{Nonlinear Systems}, 3rd ed. (Prentice Hall, 2002).




\bibitem{Lazenby2}
J. F. Lazenby, \textit{Hannibal's War: A Military History of the Second Punic War} (University of Oklahoma Press, 1978).






\end{thebibliography}

\end{document}